\documentclass[10pt]{article}
\usepackage{latexsym}
\newcommand{\be}{\begin{equation}}
\newcommand{\ee}{\end{equation}}
\def\n{\noindent}
\catcode `\@=11
\catcode `\@=12
\begin{document}
\begin{center}
\large{\bf {Cosmological Models in Lyra Geometry: Kinematics Tests}}\\
\vspace{10mm}
\normalsize{G. S. KHADEKAR} \\
\normalsize{\it Department of Mathematics, R. T. M. Nagpur University, Mahatma Jyotiba Phule
Educational Campus, Amravati Road, Nagpur - 440 033, India} \\
\normalsize{\it e-mail: gkhadekar@yahoo.com, gkhadekar@rediffmail.com} \\
\vspace{5mm}
\normalsize{ANIRUDH PRADHAN\footnote{Corresponding author} and KASHIKA SRIVASTAVA$^2$} \\
\vspace{5mm}
\normalsize{\it Department of Mathematics, Hindu Post-graduate College,
 Zamania-232 331, Ghazipur, India} \\
\normalsize{$^1$\it e-mail: pradhan@iucaa.ernet.in, acpradhan@yahoo.com}\\
\normalsize{$^2$\it e-mail: kash\_gg@yahoo.co.uk}
\end{center}
\vspace{10mm}
\begin{abstract}
In this paper the observational consequence of the cosmological
models and the expression for the neoclassical tests, luminosity
distance, angular diameter distance and look back time are
analyzed in the framework of Lyra geometry. It is interesting to
note that the space time of the universe is not only free of Big
Bang singularity but also exhibits acceleration during its
evolution.
\end{abstract}
\smallskip
\n PACS: {04.50.+h, 98.80.-k} \\
\smallskip
\n Keywords: {Cosmological tests, Singularity free model, Lyra geometry }
\section{Introduction}
Einstein geometrized gravitation. Weyl\cite{ref1} was inspired by
it and he was the first to unify gravitation and electromagnetism
in a single spacetime geometry. He showed how one can introduce a
vector field in the Riemannian spacetime with an intrinsic
geometrical significance. But this theory was not accepted as it
was based on non-integrability of length transfer. Lyra\cite{ref2}
introduced a gauge function, i.e., a displacement vector in
Riemannian spacetime which removes the non-integrability condition
of a vector under parallel transport. In this way Riemannian
geometry was given a new modification by him and the modified
geometry was named as Lyra's geometry.

Sen\cite{ref3} and Sen and Dunn\cite{ref4} proposed a new
scalar-tensor theory of gravitation and constructed the field
equations analogous to the Einstein's field equations, based on
Lyra's geometry which in normal gauge may be written in the form
\begin{equation}
\label{eq1}
R_{ij} - \frac{1}{2} g_{ij} R + \frac{3}{2} \phi_i \phi_j - \frac{3}{4} g_{ij}
\phi_k \phi^k = - 8 \pi G T_{ij},
\end{equation}
\noindent
where $\phi_{i}$ is the displacement vector and other symbols have their usual meanings.

Halford\cite{ref5} has pointed out that the constant vector displacement
field $\phi_i$ in Lyra's geometry plays the role of cosmological
constant $\Lambda$ in the normal general relativistic treatment. It
is shown by Halford\cite{ref6} that the scalar-tensor treatment based on
Lyra's geometry predicts the same effects, within observational limits
as the Einstein's theory. Several investigators{\cite{ref7}$-$\cite{ref15}} have studied
cosmological models based on Lyra geometry in different context. Soleng\cite{ref8} has
pointed out that the cosmologies based on Lyra's manifold with constant gauge vector $\phi$
will either include a creation  field and be equal to Hoyle's creation field cosmology
{\cite{ref16}$-$\cite{ref18}} or contain a special vacuum field which together with
the gauge vector term may be considered as a cosmological term. In the latter case
the solutions are equal to the general relativistic cosmologies with a cosmological term.

The purpose of this work is to analyze general features of Bianchi
type-I cosmological model with time dependent displacement vector
in the framework of Lyra geometry.
\section{Field equations}
We consider LRS Bianchi Type-I space time
\begin{equation}
\label{eq2}
ds^2 = dt^2 - A^2 dx^2 - B^2(dy^2 + dz^2),
\end{equation}
where, $A$ and $B$ are functions of $x$ and $t$. We take a perfect
fluid form for the energy momentum tensor
\begin{equation}
\label{eq3}
T_{ij} = (\rho + p) u_i u_j - p g_{ij},
\end{equation}
together with comoving coordinates $u^{i} u_{i} = 1$, where $u_{i} = (0, 0, 0, 1)$. Let
us consider a time-like displacement field vector defined by
\begin{equation}
\label{eq4}
\phi_i = (0, 0, 0, \beta(t) ).
\end{equation}
The energy momentum tensor $T^{ij}$ is not conserved in Lyra's
geometry. The essential difference between the cosmological
theories based on Lyra geometry and the Riemannian geometry lies
in the fact that the constant vector displacement field $\beta$
arises naturally from the concept of gauge in Lyra geometry
whereas the cosmological constant $\Lambda$ was introduced in {\it
adhoc} fashion in the usual treatment.

The field equations (\ref{eq1}) with the equations (\ref{eq2}),
(\ref{eq3}) and (\ref{eq4}) take the form
\begin{equation}
\label{eq5}
\frac{2\ddot B}{B} + \frac{\dot B^2}{B^2} - \frac{B^{\prime 2}}{A^2 B^2}
+ \frac{3}{4}\beta^2 = -\chi p,
\end{equation}
\begin{equation}
\label{eq6}
\dot B^{\prime} - \frac{B^{\prime} \dot A}{A} = 0,
\end{equation}
\begin{equation}
\label{eq7}
\frac{\ddot A}{A} + \frac{\ddot B}{B} + \frac{\dot A \dot B}{A B} -
\frac{B^{\prime\prime}}{A^2 B} + \frac{A^{\prime} B^{\prime}}{A^3 B}
+ \frac{3}{4}\beta^2 = - \chi p,
\end{equation}
\begin{equation}
\label{eq8}
\frac{2B^{\prime\prime}}{A^2 B} - \frac{2A^{\prime} B^{\prime}}{A^3 B}
+ \frac{B^{\prime~2}}{A^2 B^2} - \frac{2\dot A \dot B}{A B} -
\frac{\dot B^2}{B^2} + \frac{3}{4}\beta^2 = \chi \rho.
\end{equation}
The energy conservation equation is
\begin{equation}
\label{eq9} \chi \dot\rho + \frac{3}{2} \beta \dot\beta + \left[
\chi(\rho + p) + \frac{3}{2} \beta^2\right]\left(\frac{\dot A}{A}
+ 2 \frac{\dot B}{B}\right) = 0,
\end{equation}
where $\chi = 8\pi G$. Here and in what follows, a prime and a dot indicate
partial differentiation with respect to $x$ and $t$, respectively. We assume that
the fluid obeys a barotropic equation of state
\begin{equation}
\label{eq10}
p = \gamma \rho,
\end{equation}
where $\gamma (0\leq\gamma\leq 1)$ is a constant.
\section{Solutions of the field equations}
On integrating the equation (\ref{eq6}), we obtain
\begin{equation}
\label{eq11}
A = \frac{B^{\prime}}{l},
\end{equation}
where $l$ is an arbitrary function of $x$. Using equation (\ref{eq11}),
equations (\ref{eq5}) and (\ref{eq7}) can be written as
\begin{equation}
\label{eq12}
\frac{B}{B^{\prime}} \frac{d}{dx}\left(\frac{\ddot B}{B}\right) +
\frac{\dot B}{B^{\prime}}\frac{d}{dt}\left(\frac{B^{\prime}}{B}\right)
+ \frac{l^2}{B^2}\left(1 - \frac{B l^{\prime}}{B^{\prime} l}\right) = 0.
\end{equation}
Since $A$ and $B$ are separable functions of $x$, so, $\frac{B^{\prime}}{B}$
is a function of $x$. Consequently, equation (\ref{eq12}) gives after
integration
\begin{equation}
\label{eq13}
B = l S(t),
\end{equation}
where $S(t)$ is the scale factor which is an arbitrary function of $t$. Using the
equation (\ref{eq13}),
(\ref{eq11}) becomes
\begin{equation}
\label{eq14}
A = \frac{l^{\prime}}{l} S.
\end{equation}
The metric (\ref{eq5}) then takes the form
\begin{equation}
\label{eq15}
ds^2 = dt^2 - S^2 (t)\left[ dX^2 + e^{2X}(dy^2 + dz^2)\right],
\end{equation}
where $X = {\rm ln}~ l$. \\The expression for the density and
pressure from the Equations (\ref{eq8}) and (\ref{eq5}) give
\begin{equation}
\label{eq16}
\frac{3\dot{S}^{2}}{S^{2}} - \frac{3}{S^{2}} - \frac{3}{4}\beta^{2} = \rho_{d} + \rho_{x},
\end{equation}
\begin{equation}
\label{eq17}
\frac{2\ddot{S}}{S} + \frac{\dot{S}^{2}}{S^{2}} - \frac{1}{S^{2}} + \frac{3}{4}\beta^{2}
= -p_{x}.
\end{equation}
Eqs. (\ref{eq16}) and (\ref{eq17}) leads to the continuity equation
\begin{equation}
\label{eq18}
\dot{\rho}_{d} + \dot{\rho}_{x} + \frac{3}{2}\beta \dot{\beta} + 3\left[\rho_{d} + \rho_{x}
+ p_{x} + \frac{3}{2}\beta^{2}\right]\left(\frac{\dot{S}}{S}\right) = 0,
\end{equation}
where we assume $\chi = 8\pi G = 1$, $p=p_{x}$ and $\rho=
\rho_{d}+\rho_{d}$, $\rho_{d}$ is the dust matter whereas
$\rho_{x}$ refers to the missing component of the energy obeys an
equation of state
\begin{equation}
\label{eq19}
p_{x} = -\gamma \rho_{x},
\end{equation}
where $\gamma(0<\gamma<1)$ is constant.

We assume that energies densities due to dust matter and missing
component of the Universe are always less than to its critical
value $\rho_{crit} = 3(\frac{\dot{S}^{2}}{S^{2}})$. Hence one can
consider
\begin{equation}
\label{eq20}
\rho_{d} = a\rho_{crit} \; \; \; {\rm and} \; \; \;  \rho_{x} = b\rho_{crit},
\end{equation}
where the fractions $a$ (density parameter due to dust contribution), b (density parameter
due to missing mass contribution) are such that $a, b < 1$ and $a + b < 1$.

From Eqs. (\ref{eq16}) and (\ref{eq17}) by using (\ref{eq19}) we
obtain
\begin{equation}
\label{eq21}
\frac{3\ddot{S}}{S} = \frac{3\gamma - 1}{2}\rho_{x} - \frac{1}{2}\rho_{d} - \frac{3}{2} 
\beta^{2}.
\end{equation}
Now if
\begin{equation}
\label{eq22}
\frac{3\gamma - 1}{2}\rho_{x} > \frac{1}{2}\rho_{d} - \frac{3}{2} \beta^{2},
\end{equation}
then we have
$$\frac{\ddot{S}}{S} > 0.$$
This shows that the deceleration parameter
$$ q = -\frac{S\ddot{S}}{\dot{S}^{2}} < 0.$$
In other words, the Universe is accelerating because acceleration at a certain stage in the
evolution of the Universe implies $q < 0$ for some `t'.

From Eqs. (\ref{eq16}), (\ref{eq17}) and (\ref{eq20}), we get
\begin{equation}
\label{eq23}
\frac{\ddot{S}}{S} + M\frac{\dot{S}^{2}}{S^{2}} = \frac{2}{S^{2}},
\end{equation}
where
\begin{equation}    
\label{eq24}
2M = 4 - 3a -3b(1 + \gamma).
\end{equation}
Integrating Eq. (\ref{eq23}), we obtain
\begin{equation}
\label{eq25}
\int{\left[\frac{2}{M} + c_{1}S^{-2M}\right]^{-1/2}}dS = t - c_{2},
\end{equation}
where $c_{1}$ and $c_{2}$ are constants of integration.

To solve the above integration, let us choose 
$$2M = -1, i.e., 5 = 3a + 3b(1 + \gamma).$$ 
For this choice Eq. (\ref{eq25}), after integration, gives
\begin{equation}
\label{eq26}
S = \frac{4}{c_{1}} + \frac{c_{1}}{4}(t - c_{2})^{2}.
\end{equation}
Hence from the field equations the displacement vector takes the form
\begin{equation}
\label{eq27} \beta^{2} = \frac{c_{1}^{2}}{3[\frac{4}{c_{1}} +
\frac{c_{1}}{4}(t - c_{2})^{2}]^{2}}\left[\frac{1}{2}(7 - 6a -
6b)(t - c_{1})^{2} -4\right] - \frac{2c_{1}}{3[\frac{4}{c_{1}} +
\frac{c_{1}}{4}(t - c_{2})^{2}]}.
\end{equation}
The deceleration parameter $q$ and Hubble parameter $H$ are given respectively
\begin{equation}
\label{eq28}
q = -\frac{1}{2} - \frac{8}{c_{1}^{2}}\left(t - c_{2}\right),
\end{equation}
\begin{equation}
\label{eq29}
H = \frac{c_{1}}{4}(t - c_{2})\left[\frac{4}{c_{1}} + \frac{c_{1}}{4}
(t - c_{2})^{2}\right]^{-1}.
\end{equation}
In this model, the particle horizon exist because
$$\int_{-\infty}^{t}{[S(T)]^{-1}}dT$$ is a convergent integral. As
the model is singularity free, we consider $T \to -\infty$.
\section{Neoclassical Tests (Proper Distance $d(z)$)}
The proper distance between the source and observer is given by
\begin{equation}
\label{eq30}
 d(z)=S_{0}\int_{S}^{S_{0}}\frac{dS}{S \dot{S}}.
\end{equation}
From Eq. (\ref{eq26}), after integration, we get
\begin{equation}
\label{eq31}
 d(z)=S_{0}\left[\sin^{-1}{\sqrt{\frac{4(1 + z)}{c_{1}S_{0}}}} - \sin^{-1}{\sqrt{\frac{4}
{c_{1}S_{0}}}}\right],
\end{equation}
where $1 + z = \frac{S_{0}}{S}$ = redshift. This is increasing function of `z'.
\section{Luminosity Distance}

Luminosity distance is the another important concept of theoretical cosmology of a light
source. The luminosity distance is a way of expanding the amount of light received from
a distant object. It is the distance that the object appears to have, assuming the inverse
square law for the reduction of light intensity with distance holds.

If $d_{L}$ is the luminosity distance to the object, then
\begin{equation}
\label{eq32}
d_{L} = \left(\frac{L}{4\pi l}\right)^{\frac{1}{2}},
\end{equation}
where $L$ is the total energy emitted by the source per unit time,
$l$ is the apparent luminosity of the object. Therefore one can
write
\begin{equation}
\label{eq33} d_{L} = (1 + z)d(z).
\end{equation}
This is also an increasing function of `z'.
\section{Angular Diameter Distance}

The angular diameter distance is a measure of how large objects appear to be. As with
the luminosity distance, it is defined as the distance that an object of known physical
extent appears to be at, under the assumption of Euclidean geometry.

The angular diameter $d_{A}$ of a light source of proper distance $d$ is given by
\begin{equation}
\label{eq34}
d_{A} = d(z)(1+z)^{-1}= d_{L}(1 + z)^{-2},
\end{equation}
which is a decreasing function of `z'.

The angular diameter and luminosity distances have similar forms, but have a different
dependence on redshift. As with the luminosity distance, for nearly objects the angular
diameter distance closely matches the physical distance, so that objects appear smaller
as they are put further away. However the angular diameter distance has a much more
striking behaviour for distant objects. The luminosity distance effect dims the radiation
and the angular diameter distance effect means the light is spread over a large angular area.
This is so-called surface brightness dimming is therefore a particularly strong function
of redshift.
\section{Look Back Time}

The time in the past at which the light we now receive from a distant object was emitted
is called the look back time. How {\it long ago} the light was emitted (the look back time)
depends on the dynamics of the universe.

The radiation travel time (or lookback time) $(t - t_{0})$ for photon emitted by a source
at instant $t$ and received at $t_{0}$ is given by
\begin{equation}
\label{eq35} t_{0}-t  = \int^{S_{0}}_{S} \frac{dS}{\dot{S}}.
\end{equation}
By using the value from Eq. (\ref{eq26}), Eq. (\ref{eq35}) after
integration, can be expressed as
\begin{equation}
\label{eq36}
 t_{0}-t = {\frac{2\sqrt S_{0}}{c_{1}}}\left[1 - (1 +
z)^{-1/2}\right].
\end{equation}

\section {Discussions}
In the present investigation, we examine the cosmological problems
for  considering a homogeneous and isotropic cosmological model in
the framework of Lyra geometry. We have shown that space time of
the universe is free from Big Bang singularity. Moreover, the
displacement vector field $\beta$ arises in Lyra geometry, we
refer to as vacuum energy and missing energy, causes the univesre
to pass through an accelerating phase. In particular if we take $b
= 0.75$ and $\gamma = 0.95$, then `a' will be $0.204166$. This
result matches with the experiment (Today, all most all
determination of matter density are consistent with
$\rho_{d}/\rho_{crit} = a = 0.2 - 0.3$). In our model, $(a + b)
<1$, so to reach total energy density $=1$, one can assume that
the displacement vector plays the role of additional energy
density, which we have referred to as vacuum energy.

In cosmology there are many ways to specify the distance between
two points, because in the expanding universe, the distance
between comoving objects are constantly changing and Earth-bound
observers look back in time as they look out in distance. The
exact expression for  the proper distance, luminosity distance
red-shift, angular diameter distance red-shift and look back time
red-shift for the model are presented in sections $4$ to $7$, in
the framework of Lyra geometry. Beside, the implication of Lyra's
geometry for astrophysical interesting bodies is still an open
question. The problem of equations of motion and gravitational
radiation need investigation.
\section*{Acknowledgements}
\noindent The authors (G. S. Khadekar and A. Pradhan) would like to thank the 
Inter-University Center for Astronomy and Astrophysics, Pune, India for providing
facility where part of this work was carried out.
\newline
\newline

\end{document}